\documentclass[a4paper,11pt]{article}
\usepackage{a4wide,amsmath,amssymb,bbm}
\usepackage[makeroom]{cancel}
\usepackage[english]{babel}
\usepackage[utf8]{inputenc}

\usepackage{color}

\usepackage{hyperref,enumerate}
\hypersetup{
    colorlinks,
    linkcolor={black},
    citecolor={black},
    urlcolor={black}
}

\renewcommand{\o}{\overline}
\renewcommand{\u}{\underline}

\makeatletter

\@addtoreset{equation}{section} \makeatother

\begin{document}
\setcounter{page}{0}

\hfill
\vspace{30pt}

\begin{center}
{\huge{\bf {String theory at order $\alpha'^2$ and the generalized Bergshoeff-de Roo identification}}}

\vspace{80pt}

Stanislav Hronek  \ \ and \ \ Linus Wulff

\vspace{15pt}

\small {\it Department of Theoretical Physics and Astrophysics, Faculty of Science, Masaryk University\\ 611 37 Brno, Czech Republic}
\\
\vspace{12pt}
\texttt{436691@mail.muni.cz, wulff@physics.muni.cz}\\

\vspace{80pt}

{\bf Abstract}
\end{center}
\noindent
It has been shown by Marques and Nunez that the first $\alpha'$-correction to the bosonic and heterotic string can be captured in the $O(D,D)$ covariant formalism of Double Field Theory via a certain two-parameter deformation of the double Lorentz transformations. This deformation in turn leads to an infinite tower of $\alpha'$-corrections and it has been suggested that they can be captured by a generalization of the Bergshoeff-de Roo identification between Lorentz and gauge degrees of freedom in an extended DFT formalism. Here we provide strong evidence that this indeed gives the correct $\alpha'^2$-corrections to the bosonic and heterotic string by showing that it leads to a cubic Riemann term for the former but not for the latter, in agreement with the known structure of these corrections including the coefficient of Riemann cubed.

\clearpage
\tableofcontents

\section{Introduction}

Classical string theory restricted to backgrounds with $d$ abelian isometries features a continuous $O(d,d;\mathbbm R)$ symmetry \cite{Meissner:1991zj,Meissner:1991ge} which extends to all orders in $\alpha'$ \cite{Sen:1991zi}. This symmetry is closely related to T-duality. Double Field Theory (DFT) \cite{Siegel:1993th,Hull:2009mi,Hohm:2010pp} is an attempt to formulate the string effective action with an $O(D,D)$ symmetry ($D=10$ or $26$), already \emph{before} restricting to backgrounds with isometries. In order to do this one has to double the number of spacetime dimensions from $D$ to $2D$. An $O(D,D)$ invariant ``section condition'' is then imposed which reduces the number of physical coordinates down to $D$. While there is no a priori reason why this should work in general, DFT does work at the supergravity level and has proven to be a very useful tool.

Remarkably, Marques and Nunez were able to extend DFT beyond the supergravity level by showing that a certain two-parameter modification of the transformation rules leads to the first $\alpha'$-correction to the bosonic and heterotic string effective actions \cite{Marques:2015vua}, see also \cite{Bedoya:2014pma,Coimbra:2014qaa}. The modification of the transformation rules of DFT at order $\alpha'$ leads to an infinite series of $\alpha'$-corrections. In \cite{Baron:2018lve} it was argued that these can all be captured by enlarging the DFT gauge group and imposing a DFT version of a trick used by Bergshoeff and de-Roo \cite{Bergshoeff:1988nn} to find $\alpha'$-corrections to the heterotic string.\footnote{This idea was previously used in \cite{Lee:2015kba} to find the heterotic $\alpha'^2$-correction in DFT.} They dubbed this idea the ``generalized Bergshoeff de-Roo identification''. The construction is somewhat formal since it requires and infinite-dimensional gauge group, but nevertheless, the identification can be solved recursively order-by-order in $\alpha'$, leading to specific corrections to the DFT action and transformation rules at each order. Unfortunately, the expressions found at order $\alpha'^2$ in \cite{Baron:2020xel} take a very complicated form and it was not possible to compare them to the known corrections to the bosonic and heterotic string.

Here we will show that many of the terms in the $\alpha'^2$ corrected DFT action of \cite{Baron:2020xel} are zero due to Bianchi identities, or can be removed by field redefinitions. In this way we are able to show that their expressions give rise to a cubic Riemann term, plus quartic terms which we don't determine, in the bosonic string case and no cubic terms in the heterotic case. This is in precise agreement with the known structure of the $\alpha'^2$-correction to the bosonic and heterotic string (in the NS sector) \cite{Metsaev:1986yb}, including the coefficient of the cubic Riemann term.\footnote{The first correction to the type II string is at order $\alpha'^3$.}

We should note that the fact that the DFT formalism seems to correctly capture all the corrections up to order $\alpha'^2$ does not mean it can capture all $\alpha'$-corrections. In fact, at order $\alpha'^3$ all string theories have quartic Riemann terms with coefficient $\zeta(3)$. These cannot be accounted for by the construction of \cite{Baron:2018lve}, due to the $\zeta(3)$ coefficient. In fact, a careful analysis \cite{Hronek:2020xxi} shows that they cannot be captured by the DFT formalism at all (at least not without some drastic modification of the formalism). Therefore it seems like the DFT formalism can account for the all-order T-duality completion of the Riemann squared correction at order $\alpha'$ (which in turn, in the heterotic case, is needed for the Green-Schwarz anomaly cancellation mechanism \cite{Green:1984sg}), but not for other $\alpha'$-corrections. It may seem strange that DFT can account for any $\alpha'$-corrections at all, but in our point of view this is because of the existence of the generalized Bergshoeff-de Roo identification which allows the $\alpha'$-corrections connected with Riemann squared to be generated for free from an uncorrected (extended) DFT action.

The rest of this note is organized as follows. In section \ref{sec:DFT} we give a short summary of the elements of the flux formulation of DFT which we will need. Then we describe the main steps in the calculations at order $\alpha'^2$ in section \ref{sec:calculation}. We end with some conclusions. Details of the calculations are provided in the appendix.

\section{\boldmath Elements of the \texorpdfstring{$O(D,D)$}{O(D,D)} covariant formulation}\label{sec:DFT}
Here we will introduce the elements of the $O(D,D)$ covariant formulation of DFT which we will need. We will use the so-called flux formulation of \cite{Geissbuhler:2013uka,Marques:2015vua}, see also \cite{Lescano:2021lup} for a recent review. 

The basic object is the generalized vielbein which we parametrize as
\begin{equation}
E_A{}^M=
\frac{1}{\sqrt2}
\left(
\begin{array}{cc}
e^{(+)a}{}_m-e^{(+)an}B_{nm} & e^{(+)am}\\
-e^{(-)}_{am}-e^{(-)}_a{}^nB_{nm} & e^{(-)}_a{}^m
\end{array}
\right)\,.
\label{eq:E}
\end{equation}
The two sets of vielbeins $e^{(\pm)}$ for the metric $G_{mn}$ transform independently as $\Lambda^{(\pm)}e^{(\pm)}$ under two copies of the Lorentz-group. The standard supergravity fields are recovered by fixing the gauge $e^{(+)}=e^{(-)}=e$, leaving only the diagonal copy of the Lorentz-group. The dilaton $\Phi$ is encoded in the generalized dilaton $d$ defined as
\begin{equation}
e^{-2d}=e^{-2\Phi}\sqrt{-G}\,.
\end{equation}
There are two constant metrics, the $O(D,D)$ metric $\eta^{AB}$ and the generalized metric $\mathcal H^{AB}$, which take the form
\begin{equation}
\eta^{AB}=
\left(
\begin{array}{cc}
	\hat\eta_{ab} & 0\\
	0 & -\hat\eta^{ab}
\end{array}
\right)\,,\qquad
\mathcal H^{AB}=
\left(
\begin{array}{cc}
	\hat\eta_{ab} & 0\\
	0 & \hat\eta^{ab}
\end{array}
\right)\,,
\label{eq:eta}
\end{equation}
where $\hat\eta=(-1,1,\ldots,1)$ is the $D$-dimensional Minkowski metric. The $O(D,D)$-metric is used to raise/lower indices. From these we build the projection operators
\begin{equation}
P_\pm^{AB}=\frac12\left(\eta^{AB}\pm\mathcal H^{AB}\right)\,.
\label{eq:Ppm}
\end{equation}
We denote projected indices by over/underlining them and use lower-case letters for non-doubled indices, e.g.\footnote{
In expressions where the indices are suppressed we will use the notation
$$
F^{(\pm)}_{ABC}=(P_{\mp})_A{}^D(P_{\pm})_B{}^E(P_{\pm})_C{}^F F_{DEF}\,,\qquad
 F^{(\pm\pm)}_{ABC}=(P_{\pm})_A{}^D(P_{\pm})_B{}^E(P_{\pm})_C{}^F F_{DEF}\,.
$$
}
\begin{equation}
P^{AB}_{+}F_B\rightarrow F^{\o a}\,,\qquad P^{AB}_{-}F_B\rightarrow F^{\u a}\,.
\end{equation}
We define the derivative with a ``flat'' index as
\begin{equation}
\partial_A=E_A{}^M\partial_M\,,
\end{equation}
where the standard solution to the section condition is $\partial_M=(0,\partial_m)$.

The diffeomorphism and B-field gauge transformation invariant information in the generalized vielbein is encoded in the basic generalized diffeomorphism scalars
\begin{equation}
F_{ABC}=3\partial_{[A}E_B{}^ME_{C]M}\,,\qquad F_A=\partial^BE_B{}^ME_{AM}+2\partial_Ad\,.
\label{eq:fluxes}
\end{equation}
These ``generalized fluxes'' are manifestly $O(D,D)$ invariant and they are the basic building blocks from which to construct an $O(D,D)$ invariant action. Indeed, the lowest order action takes the form
\begin{equation}
S=\int dX\,e^{-2d}\mathcal R\,,
\label{eq:S0}
\end{equation}
where the generalized Ricci scalar is defined as
\begin{equation}
\mathcal R
=
4\partial^{\o a}F_{\o a}
-2F^{\o a}F_{\o a}
+F_{\u a\o{bc}}F^{\u a\o{bc}}
+\frac13F_{\o{abc}}F^{\o{abc}}\,.
\label{eq:R}
\end{equation}
By gauge-fixing $e^{(+)}=e^{(-)}=e$ and imposing the standard solution to the section condition,$\partial_M=(0,\partial_m)$, one recovers the low-energy effective action of bosonic string theory, or the NS sector of the heterotic string. Note that $O(D,D)$ symmetry and generalized diffeomorphism symmetry are manifest in this formulation, but the (doubled) Lorentz symmetry is not manifest and must be checked by hand.

A sequence of higher order $\alpha'$-corrections to the action (\ref{eq:S0}) can be derived using the ``generalized Bergshoeff-de Roo identification'', i.e. extending the duality group and double Lorentz group and then identifying the new gauge vectors with the generalized spin connection, as described in \cite{Baron:2020xel}. To the order we are interested in here it takes the form
\begin{equation}
S=\int dX\,e^{-2d}\left(\mathcal R^{(0,0)}+a\mathcal  R^{(0,1)}+b\mathcal R^{(1,0)}+ a^2\mathcal{R}^{(0,2)}+ab\mathcal{R}^{(1,1)}+b^2\mathcal{R}^{(2,0)}\right)\,.
\label{eq:S-DFT}
\end{equation}
Here $\mathcal R^{(0,0)}=\mathcal R$ gives the lowest order action and $a,b$ are parameters proportional to $\alpha'$. The bosonic string result is recovered by taking $a=b=-\alpha'$ and the heterotic string result by taking $a=-\alpha'$ and $b=0$. At the first order in $\alpha'$ we have \cite{Hronek:2020skb}
\begin{align}
\mathcal R^{(0,1)}
=&
-(\partial^{\o a}-F^{\o a})\left[(\partial^{\o b}-F^{\o b})\big(F_{\o a\u{cd}}F_{\o b}{}^{\u{cd}}\big)\right]
-\tfrac12\mathcal R_{\o{ab}\u{cd}}\mathcal R^{\o{ab}\u{cd}}
+\partial^{\o a}F^{\o b}F_{\o a\u{cd}}F_{\o b}{}^{\u{cd}}
\\
&{}
+F^{\o{ab}C}F_{\o a}{}^{\u{de}}\partial_CF_{\o b\u{de}}
-\tfrac{2}{3}F^{\o{abc}}F_{\o a\u d}{}^{\u e}F_{\o b\u e}{}^{\u f}F_{\o c\u f}{}^{\u d}
+\left(F^{\u a\o{bc}}F_{\u a\o{bd}}+\tfrac12F^{\o{abc}}F_{\o{abd}}\right)F_{\o c\u{ef}}F^{\o d\u{ef}}\,,
\nonumber
\end{align}
while $\mathcal R^{(1,0)}$ takes the same form, but with over and underlined indices exchanged. The first term is a total derivative, while in the second term we have introduced the ``generalized Riemann tensor''\footnote{Reversing the projections we get the same object up to a sign, $\mathcal R_{\u{cd}\o{ab}}=-\mathcal R_{\o{ab}\u{cd}}$, due to Bianchi identities.}
\begin{equation}
\mathcal R_{\o{ab}\u{cd}}
=
2\partial_{[\o a}F_{\o b]\u{cd}}
-F_{\o{abe}}F^{\o e}{}_{\u{cd}}
-2F_{[\o a|\u c|}{}^{\u e}F_{\o b]\u{ed}}\,.
\end{equation}
The quotation marks are there to emphasize that unlike the usual Riemann tensor this object does not transform covariantly under double Lorentz transformations. Indeed, going to the usual supergravity fields one finds
\begin{equation}
\mathcal R_{\o{ab}\u{cd}}\rightarrow\frac12\left(R^{(-)ab}{}_{cd}+\omega^{(+)eab}\omega^{(-)}_{ecd}\right)\,,
\label{eq:R-}
\end{equation}
where we have defined the torsionful spin connections $\omega^{(\pm)}=\omega\pm\frac12H$ and $R^{(-)}$ is the curvature of $\omega^{(-)}$. This $\alpha'$-correction to the action agrees with the more complicated original expression found in \cite{Marques:2015vua}, which was shown there to reproduce the known $\alpha'$-correction to the bosonic and heterotic string.

Expressions for $\mathcal{R}^{(0,2)}$, $\mathcal{R}^{(1,1)}$ and $\mathcal{R}^{(2,0)}$ were found in \cite{Baron:2020xel}, but the expressions are very long indeed. Only $\mathcal{R}^{(0,2)}$ and $\mathcal{R}^{(1,1)}$ are relevant since $\mathcal{R}^{(2,0)}$ is simply related to $\mathcal{R}^{(0,2)}$ by reversing all projections. They consist of about 280 and 190 terms respectively! For this reason it is very difficult to compare the expressions of \cite{Baron:2020xel} with the known $\alpha'^2$-correction to the bosonic and heterotic string. However, we will show here that many terms in these complicated expressions vanish upon field redefinitions and using Bianchi identities so that one eventually finds
\begin{equation}
\mathcal{R}^{(0,2)}\sim \mathcal{R}^{(2,0)}\sim\mathcal O(F^4)\,,\qquad \mathcal{R}^{(1,1)}\sim-\frac13\mathcal R_{\o a}{}^{\o b\u{de}}\mathcal R_{\o{bc}\u{ef}}\mathcal R^{\o{ca}\u f}{}_{\u d}+\mathcal O(F^4)\,.
\end{equation}
Since only $\mathcal{R}^{(0,2)}$ enters in the heterotic case one finds no cubic Riemann terms in that case. This is in agreement with scattering amplitude calculations \cite{Metsaev:1986yb}. For the bosonic string on the other hand we get a cubic Riemann term from $\mathcal{R}^{(1,1)}$, again in agreement with the known structure of the $\alpha'^2$-correction to the bosonic string, including the coefficient \cite{Metsaev:1986yb,Garousi:2019mca}.\footnote{The $\nabla H\nabla HR$-terms can be removed by field redefinitions.} This provides strong evidence that the expressions derived using the generalized Bergshoeff-de Roo identification indeed reproduce also the $\alpha'^2$-corrections to the bosonic and heterotic string. 

Before we describe the calculations, we give a summary of the identities needed.

\subsection{Useful identities}
The following identities are used throughout the calculation:
\begin{itemize}
    \item The section condition 
\begin{equation}
\label{sc}
  \partial_AY\partial^AZ=0\,.
\end{equation}    
    \item Integration by parts
\begin{equation}
    \int e^{-2d}\partial_AYZ =-\int e^{-2d}Y\left(\partial_{A}-F_{A}\right)Z\,.
    \label{ibp}
\end{equation}
    \item Commutation of derivatives
\begin{equation}
    [\partial_A,\partial_B]=F_{ABC}\,\partial^C.  
\label{der}
\end{equation}
    \item Bianchi identities
    \begin{equation}
4\partial_{[A}F_{BCD]}=3F_{[AB}{}^EF_{CD]E}\,,\qquad
2\partial_{[A}F_{B]}=-(\partial^C-F^C)F_{ABC}\,.
\label{eq:Bianchi}
\end{equation}
\item Equations of motion (terms involving these are removed by field redefinitions)\footnote{The same equations with over and underlined indices exchanged also hold.}
\begin{align}
4\partial^{\o a}F_{\o a}
-2F^{\o a}F_{\o a}
+F_{\u a\o{bc}}F^{\u a\o{bc}}
+\frac13F_{\o{abc}}F^{\o{abc}}=&0\,,\\
\partial^{\o{a}}F^{\u{b}}+(\partial^{\u{c}}-F^{\u{c}})F^{\o{a}\u{b}}{}_{\u{c}}-F^{\u{c}\o{da}}F_{\o{d}\u{c}}{}^{\u{b}}=&0\,.
\label{eq:eom}
\end{align}
\end{itemize}
It follows from these that, up to equations of motion terms,
\begin{equation}
    \partial^{\o{a}}\mathcal R_{\o{ab}\u{cd}}\sim\mathcal O(F^2)\,,\quad \partial^{\o{a}}\partial_{\o{a}}\mathcal R_{\o{bc}\u{de}}\sim\mathcal O(F^2)\,,
\label{cur}    
\end{equation}
or, equivalently, 
\begin{equation}
    \partial^{\o{a}}\partial_{\o{a}}F^{\o{b}\u{cd}}\sim\partial_{\o{a}}\partial^{\o{b}}F^{\o{a}\u{cd}}+\mathcal O(F^2)\,.
\label{id}
\end{equation}

\section{Simplification of \texorpdfstring{$\alpha'^2$}{alpha'**2} terms}\label{sec:calculation}
Here we will describe the main steps in our calculations. We will use the expressions given in \cite{Baron:2020xel}, but it is important to remember that their action differs from ours by an overall factor of 2. Therefore, to get the result in our conventions, we have to multiply their expressions by 2. The details are relegated to the appendix.

\subsection{\texorpdfstring{$\mathcal{R}^{(0,2)}$}{R(0,2)} term}
It is useful to split the complicated formula for $\mathcal{R}^{(0,2)}$ from \cite{Baron:2020xel} into a number of pieces. Firstly we split it into $\mathcal{R}^{(0,2)}_{\Phi}$ and $\mathcal{R}^{(0,2)}_{\cancel{\Phi}}$ according to the dilaton dependence which is encoded in the one index flux $F_A$. Next we use a superscript in parenthesis to denote the powers of $F$ involved, dropping all terms of order $F^4$ and higher. Lastly, we use a subscript to distinguish different types of terms mostly according to the projections involved.

\noindent
Terms of second order in fields with no $F_A$:
\begin{align}
    \left[\mathcal R^{(0,2)}_{\cancel{\Phi}}\right]^{(2)}
    =&
    -\frac{1}{2}\partial^{\u{c}}\partial^{\u{d}}F^{\o{b}\u{gh}}\partial_{\u{c}}\partial_{\u{d}}F_{\o{b}\u{gh}}
    +\partial^{\overline{c}}\partial^{\underline{b}}F_{\overline{c}}{}^{\underline{ef}}\partial^{\overline{d}}\partial_{\underline{b}}F_{\overline{d}\underline{ef}}
    +\frac{1}{2}\partial^{\overline{b}}\partial^{\underline{b}}F^{\overline{d}\underline{ef}}\partial_{\overline{b}}\partial_{\underline{b}}F_{\overline{d}\underline{ef}}
    \nonumber\\
    &{}
+\frac{1}{2}\partial^{\underline{b}}F^{\overline{b}\underline{ef}}\partial^{\overline{d}}\partial_{\overline{b}}\partial_{\underline{b}}F_{\overline{d}\underline{ef}}
+\frac{3}{2}\partial^{\underline{b}}F^{\overline{a}\underline{ef}}\partial_{\overline{a}}\partial^{\overline{d}}\partial_{\underline{b}}F_{\overline{d}\underline{ef}}\,.
    \label{R2}
\end{align}
Terms of third order in fields containing $F_A$:
\begin{align}
    \label{R3withf}
    &\left[\mathcal R^{(0,2)}_{\Phi}\right]^{(3)}
    =
-2\partial^{\u{b}}F^{\o{d}\u{ef}}\partial^{\o{c}}\partial_{\u{b}}F_{\o{c}\u{ef}}F_{\o{d}}
-2\partial^{\u{b}}F^{\o{c}\u{ef}}\partial_{\o{c}}\partial_{\u{b}}F^{\o{d}}{}_{\u{ef}}F_{\o{d}}
-2\partial^{\u{b}}F^{\o{c}\u{ef}}\partial_{\u{b}}F^{\o{d}}{}_{\u{ef}}\partial_{\o{c}}F_{\o{d}}\,.
\end{align}
Terms of third order in fields not containing $F_A$ and with the following projections (up to the section condition) $\partial^{(+)}\partial^{(+)}\partial^{(-)}F^{(-)}F^{(-)}F^{(+)}$:
\begin{align}
\label{R31}
&\left[\mathcal R^{(0,2)}_{\cancel{\Phi}}\right]^{(3)}_{1}
=
-\frac{1}{2}
\partial^{\o{c}}\partial^{\o{e}}\partial^{\u{f}}F_{\o{c}}{}^{\u{de}}F^{\o{f}}{}_{\u{de}}F_{\o{ef}\u{f}}
-\frac{3}{2}
\partial^{\o{e}}\partial^{\o{c}}\partial^{\u{f}}F_{\o{c}}{}^{\u{de}}F^{\o{f}}{}_{\u{de}}F_{\o{ef}\u{f}}
+\frac{3}{2}
\partial^{\u{h}}\partial^{\u{c}}F^{\o{c}\u{fg}}\partial_{\u{c}}F^{\o{d}}{}_{\u{fg}}F_{\o{cd}\u{h}}
\nonumber\\
&{}
-2
\partial^{\o{c}}\partial^{\u{d}}F_{\o{c}}{}^{\u{ef}}\partial^{\o{d}}F_{\o{d}}{}^{\o{f}}{}_{\u{d}}F_{\o{f}\u{ef}}
-
\partial^{\o{b}}\partial^{\u{d}}F^{\o{d}\u{ef}}\partial_{\o{b}}F_{\o{d}}{}^{\o{f}}{}_{\u{d}}F_{\o{f}\u{ef}}
-2
\partial^{\o{c}}\partial^{\u{f}}F_{\o{c}}{}^{\u{de}}\partial^{\o{e}}F^{\o{f}}{}_{\u{de}}F_{\o{ef}\u{f}}
+\partial^{\u{c}}\partial^{\u{f}}F^{\o{b}\u{gh}}\partial_{\u{c}}F_{\o{b}}{}^{\o{d}}{}_{\u{f}}F_{\o{d}\u{gh}}
\nonumber\\
&{}
+
\partial^{\u{c}}\partial^{\u{h}}F^{\o{c}\u{fg}}\partial_{\u{c}}F^{\o{d}}{}_{\u{fg}}F_{\o{cd}\u{h}}
-
\partial^{\o{b}}\partial^{\u{f}}F^{\o{e}\u{de}}\partial_{\o{b}}F^{\o{f}}{}_{\u{de}}F_{\o{ef}\u{f}}
-\frac{1}{2}
\partial^{\u{d}}F^{\o{b}\u{ef}}\partial^{\o{d}}\partial_{\o{b}}F_{\o{d}}{}^{\o{f}}{}_{\u{d}}F_{\o{f}\u{ef}}
-\frac{3}{2}
\partial^{\u{d}}F^{\o{a}\u{ef}}\partial_{\o{a}}\partial^{\o{d}}F_{\o{d}}{}^{\o{f}}{}_{\u{d}}F_{\o{f}\u{ef}}
\nonumber\\
&{}
-\frac{1}{2}
\partial^{\u{f}}F^{\o{b}\u{de}}\partial^{\o{e}}\partial_{\o{b}}F^{\o{f}}{}_{\u{de}}F_{\o{ef}\u{f}}
-\frac{3}{2}
\partial^{\u{f}}F^{\o{a}\u{de}}\partial_{\o{a}}\partial^{\o{e}}F^{\o{f}}{}_{\u{de}}F_{\o{ef}\u{f}}
-2
\partial^{\u{f}}F^{\o{b}\u{de}}\partial^{\o{e}}F^{\o{f}}{}_{\u{de}}\partial_{\o{b}}F_{\o{ef}\u{f}}
\nonumber\\
&{}
+2
\partial^{\u{f}}F^{\o{a}\u{de}}\partial_{\o{a}}F^{\o{e}}{}_{\u{de}}\partial^{\o{f}}F_{\o{ef}\u{f}}\,.
\end{align}
Terms of third order in fields not containing $F_A$ and with the following projections (up to the section condition) $\partial^{(+)}\partial^{(+)}\partial^{(+)}F^{(-)}F^{(-)}F^{(-)}$:
\begin{align}
    \label{R32}
&\left[\mathcal R^{(0,2)}_{\cancel{\Phi}}\right]^{(3)}_{2}
=
-\frac{1}{2}\partial^{\o{c}}\partial^{\o{e}}\partial_{\o{c}}F^{\o{f}\u{ce}}F_{\o{e}\u{c}}{}^{\u{f}}F_{\o{f}\u{ef}}+\partial^{\o{d}}\partial^{\o{e}}\partial^{\o{f}}F_{\o{d}}{}^{\u{ce}}F_{\o{e}\u{c}}{}^{\u{f}}F_{\o{f}\u{ef}}
+3\partial^{\o{e}}\partial^{\o{d}}\partial^{\o{f}}F_{\o{d}}{}^{\u{ce}}F_{\o{e}\u{c}}{}^{\u{f}}F_{\o{f}\u{ef}}
\nonumber\\
&{}
-\frac{3}{2}\partial^{\o{e}}\partial^{\o{c}}\partial_{\o{c}}F^{\o{f}\u{ce}}F_{\o{e}\u{c}}{}^{\u{f}}F_{\o{f}\u{ef}}
+\frac{3}{2}
\partial^{\o{c}}\partial^{\o{f}}F^{\o{e}\u{ce}}\partial_{\o{c}}F_{\o{e}\u{c}}{}^{\u{f}}F_{\o{f}\u{ef}}
-
\partial^{\o{c}}\partial^{\o{e}}F^{\o{f}\u{ce}}\partial_{\o{c}}F_{\o{e}\u{c}}{}^{\u{f}}F_{\o{f}\u{ef}}
+\frac{3}{2}
\partial^{\o{c}}\partial^{\o{d}}F_{\o{d}}{}^{\u{ce}}\partial_{\o{c}}F^{\o{f}}{}_{\u{c}}{}^{\u{f}}F_{\o{f}\u{ef}}
\nonumber\\
&{}
+4
\partial^{\o{d}}\partial^{\o{f}}F_{\o{d}}{}^{\u{ce}}\partial^{\o{e}}F_{\o{e}\u{c}}{}^{\u{f}}F_{\o{f}\u{ef}}
-5
\partial^{\o{d}}\partial^{\o{e}}F_{\o{d}}{}^{\u{ce}}\partial^{\o{f}}F_{\o{e}\u{c}}{}^{\u{f}}F_{\o{f}\u{ef}}
+\frac{1}{2}
\partial^{\o{d}}\partial^{\o{c}}F_{\o{d}}{}^{\u{ce}}\partial_{\o{c}}F^{\o{f}}{}_{\u{c}}{}^{\u{f}}F_{\o{f}\u{ef}}
-3
\partial^{\o{e}}\partial^{\o{d}}F_{\o{d}}{}^{\u{ce}}\partial^{\o{f}}F_{\o{e}\u{c}}{}^{\u{f}}F_{\o{f}\u{ef}}
\nonumber\\
&{}
+\frac{1}{2}
\partial^{\o{f}}\partial^{\o{c}}F^{\o{e}\u{ce}}\partial_{\o{c}}F_{\o{e}\u{c}}{}^{\u{f}}F_{\o{f}\u{ef}}
-2
\partial^{\o{b}}\partial_{\o{b}}F^{\o{f}\u{ce}}\partial^{\o{e}}F_{\o{e}\u{c}}{}^{\u{f}}F_{\o{f}\u{ef}}
+2
\partial^{\o{b}}\partial_{\o{b}}F^{\o{e}\u{ce}}\partial^{\o{f}}F_{\o{e}\u{c}}{}^{\u{f}}F_{\o{f}\u{ef}}
+
\partial^{\u{b}}\partial^{\o{c}}F^{\o{d}\u{eg}}\partial_{\u{b}}F_{\o{c}\u{e}}{}^{\u{h}}F_{\o{d}\u{gh}}
\nonumber\\
&{}
-2
\partial^{\u{b}}\partial^{\o{d}}F^{\o{c}\u{eg}}\partial_{\u{b}}F_{\o{c}\u{e}}{}^{\u{h}}F_{\o{d}\u{gh}}
-3
\partial^{\o{b}}F^{\o{c}\u{ce}}\partial_{\o{c}}F_{\o{f}\u{ef}}\partial_{\o{b}}F^{\o{f}}{}_{\u{c}}{}^{\u{f}}
+4
\partial^{\u{b}}F^{\o{c}\u{eg}}\partial^{\o{d}}\partial_{\u{b}}F_{\o{c}\u{e}}{}^{\u{h}}F_{\o{d}\u{gh}}
-4
\partial^{\u{b}}F^{\o{a}\u{eg}}\partial_{\o{a}}\partial_{\u{b}}F_{}^{\o{d}}{}_{\u{e}}{}^{\u{h}}F_{\o{d}\u{gh}}
\nonumber\\
&{}
+7
\partial^{\u{b}}F^{\o{a}\u{eg}}\partial_{\u{b}}F^{\o{d}}{}_{\u{e}}{}^{\u{h}}\partial_{\o{a}}F_{\o{d}\u{gh}}\,.
\end{align}
Terms of the same type as the previous ones, but which trivially cancel among themselves:
\begin{align}
\label{R33}
\left[\mathcal R^{(0,2)}_{\cancel{\Phi}}\right]^{(3)}_{3}
=&
\partial^{\o{e}}\partial^{\o{d}}F^{\o{c}\u{ce}}\partial_{\o{c}}F_{\o{e}\u{c}}{}^{\u{f}}F_{\o{f}\u{ef}}
-
\partial^{\o{d}}\partial^{\o{e}}F^{\o{c}\u{ce}}\partial_{\o{c}}F_{\o{e}\u{c}}{}^{\u{f}}F_{\o{f}\u{ef}}
+2
\partial^{\u{g}}F^{\o{c}\u{ef}}\partial^{\o{d}}\partial^{\u{h}}F_{\o{c}\u{ef}}F_{\o{d}\u{gh}}
\nonumber\\
&{}
-
\partial^{\u{g}}F^{\o{a}\u{ef}}\partial_{\o{a}}\partial^{\u{h}}F^{\o{d}}{}_{\u{ef}}F_{\o{d}\u{gh}}
+2
\partial^{\u{g}}F^{\o{a}\u{ef}}\partial^{\u{h}}F^{\o{d}}{}_{\u{ef}}\partial_{\o{a}}F_{\o{d}\u{gh}}\,.
\end{align}
The only term with an $F^{(++)}$ projection:
\begin{align}
\label{R34}
&\left[\mathcal R^{(0,2)}_{\cancel{\Phi}}\right]^{(3)}_{4}
=
-\frac{1}{2}
\partial^{\u{b}}F^{\o{d}\u{ef}}\partial^{\o{e}}\partial_{\u{b}}F^{\o{f}}{}_{\u{ef}}F_{\o{def}}\,.
\end{align}

Using the identities from section \ref{sec:DFT} one finds that the quadratic terms, $\left[\mathcal R^{(0,2)}_{\cancel{\Phi}}\right]^{(2)}$, reduce to cubic terms upon suitable field redefinitions and integration by parts
\begin{align}
    \label{R201}
&
\left[\mathcal R^{(0,2)}_{\cancel{\Phi}}\right]^{(2)}
\sim
-\partial^{\underline{d}}F^{\overline{b}\underline{gh}}F^{\underline{c}}\partial_{\underline{c}}\partial_{\underline{d}}F_{\overline{b}\underline{gh}}
+
\partial^{\underline{b}}F_{\overline{b}}{}^{\underline{ef}}F^{\overline{b}}\partial^{\overline{d}}\partial_{\underline{b}}F_{\overline{d}\underline{ef}}
+
\frac{1}{2}\partial^{\underline{b}}F^{\overline{b}\underline{ef}}F_{\o{b}}{}^{\o{d}A}\partial_A\partial_{\underline{b}}F_{\overline{d}\underline{ef}}
\nonumber\\
&{}
+
\partial^{\underline{b}}F^{\overline{b}\underline{ef}}\partial^{\overline{d}}\left(F_{\o{b}\u{b}}{}^{A}\partial_AF_{\overline{d}\underline{ef}}\right)
+
3\partial^{\underline{b}}F^{\overline{b}\underline{ef}}\partial^{\overline{d}}\partial_{\underline{b}}\left(F_{[\overline{bd}}{}^EF_{\underline{ef}]E}\right)
+
\partial^{\underline{b}}F^{\overline{b}\underline{ef}}\partial^{\overline{d}}\left(F_{\underline{b}\overline{d}A}\partial^AF_{\overline{b}\underline{ef}}\right)
\nonumber\\
&{}
+
\partial^{\underline{b}}F^{\overline{b}\underline{ef}}F^{\overline{d}}{}_{\underline{b}}{}^{A}\partial_A\mathcal R_{\o{bd}\u{ef}}
-2
\partial^{\underline{b}}F^{\overline{b}\underline{ef}}\partial_{\underline{b}}\left(F^{\overline{d}}{}_{\underline{e}}{}^{A}\partial_{A}F_{\underline{f}\overline{bd}}\right)
+
\partial^{\underline{b}}F^{\overline{b}\underline{ef}}\partial_{\underline{b}}\left(F_{\underline{ef}A}\partial^AF_{\overline{b}}\right)
\nonumber\\
&{}
-2
\partial^{\underline{b}}F^{\overline{b}\underline{ef}}\partial_{\underline{b}}\partial_{\underline{e}}\left(
F^{\o{c}}F_{\u{f}\o{bc}}+F_{\o{c}\u{df}}F^{\underline{d}\o{c}}{}_{\o{b}}
\right)
+\mathcal O(F^4)\,.
\end{align}
While for the cubic terms one finds
\begin{align}
\left[\mathcal R^{(0,2)}_{\Phi}\right]^{(3)}\sim \left[\mathcal R^{(0,2)}_{\cancel{\Phi}}\right]^{(3)}_{2}\sim&\left[\mathcal R^{(0,2)}_{\cancel{\Phi}}\right]^{(3)}_{3}\sim\mathcal O(F^4)\,,
\quad
\left[\mathcal R^{(0,2)}_{\cancel{\Phi}}\right]^{(3)}_{1}\sim\frac{3}{2}\partial^{\u{h}}\partial^{\u{c}}F^{\o{c}\u{fg}}\partial_{\u{c}}F^{\o{d}}{}_{\u{fg}}F_{\o{cd}\u{h}}+\mathcal O(F^4)\,.
\end{align}
Finally, one finds that
\begin{align}
\mathcal R^{(0,2)}=&\,\left[\mathcal R^{(0,2)}_{\cancel{\Phi}}\right]^{(2)}
+\left[\mathcal R^{(0,2)}_{\Phi}\right]^{(3)}
+\left[\mathcal R^{(0,2)}_{\cancel{\Phi}}\right]^{(3)}_{1}
+\left[\mathcal R^{(0,2)}_{\cancel{\Phi}}\right]^{(3)}_{2}
+\left[\mathcal R^{(0,2)}_{\cancel{\Phi}}\right]^{(3)}_{3}
+\left[\mathcal R^{(0,2)}_{\cancel{\Phi}}\right]^{(3)}_{4}
+\mathcal O(F^4)
\nonumber\\
\sim&\,\mathcal O(F^4)\,.
\end{align}

\subsection{\texorpdfstring{$\mathcal{R}^{(1,1)}$}{R(1,1)} term}
We do a similar splitting for the $\mathcal{R}^{(1,1)}$ terms from \cite{Baron:2020xel}. There are no terms quadratic in fields.

\noindent
Terms of third order in fields without $F_A$ with projections of the type $\partial^{(+)}\partial^{(+)}\partial^{(+)}$

\noindent
$\cdot F^{(-)}F^{(-)}F^{(-)}$:
\begin{align}
    \left[\mathcal{R}^{(1,1)}_{\cancel{\Phi}}\right]^{(3)}_{1}
    =&
    \frac{4}{3}
\partial^{\o{e}}F^{\o{c}\u{ce}}\partial_{\o{c}}F_{\o{f}\u{ef}}\partial^{\o{f}}F_{\o{e}\u{c}}{}^{\u{f}}
-4
\partial^{\o{b}}F^{\o{c}\u{ce}}\partial_{\o{c}}F^{\o{f}}{}_{\u{ef}}\partial_{\o{b}}F_{\o{f}\u{c}}{}^{\u{f}}\,.
\label{eq:R11-1}
\end{align}
Terms of third order in fields without $F_A$ with projections of the type $\partial^{(+)}\partial^{(+)}\partial^{(-)}$

\noindent $\cdot F^{(-)}F^{(-)}F^{(+)}$:
\begin{align}
    &\left[\mathcal{R}^{(1,1)}_{\cancel{\Phi}}\right]^{(3)}_{2}
    =
\frac{1}{2}
\partial^{\u{f}}\partial^{\o{d}}\partial^{\o{e}}F^{\o{f}\u{de}}F_{\o{d}\u{de}}F_{\o{ef}\u{f}}
+\frac{1}{2}
\partial^{\o{d}}\partial^{\u{f}}\partial^{\o{e}}F^{\o{f}\u{de}}F_{\o{d}\u{de}}F_{\o{ef}\u{f}}
+
\partial^{\o{f}}\partial^{\u{b}}F^{\o{be}}{}_{\u{b}}\partial_{\o{b}}F_{\o{e}}{}^{\u{ef}}F_{\o{f}\u{ef}}
\nonumber\\
&{}
+\frac{1}{2}
\partial^{\u{d}}\partial^{\o{d}}F^{\o{e}\u{ef}}\partial^{\o{f}}F_{\o{de}\u{d}}F_{\o{f}\u{ef}}
+
\partial^{\u{f}}\partial^{\o{c}}F_{\o{c}}{}^{\u{de}}\partial^{\o{e}}F^{\o{f}}{}_{\u{de}}F_{\o{ef}\u{f}}
+\partial^{\u{f}}\partial^{\o{e}}F^{\o{f}\u{de}}\partial^{\o{d}}F_{\o{d}\u{de}}F_{\o{ef}\u{f}}
+
\partial^{\u{b}}F^{\o{be}}{}_{\u{b}}\partial^{\o{f}}\partial_{\o{b}}F_{\o{e}}{}^{\u{ef}}F_{\o{f}\u{ef}}
\nonumber\\
&{}
+
\partial^{\u{b}}F^{\o{ae}}{}_{\u{b}}\partial_{\o{a}}F_{\o{e}}{}^{\u{ef}}\partial^{\o{f}}F_{\o{f}\u{ef}}
+
\frac{1}{2}
\partial^{\u{f}}F^{\o{a}\u{de}}\partial_{\o{a}}\partial^{\o{e}}F^{\o{f}}{}_{\u{de}}F_{\o{ef}\u{f}}
+\partial^{\u{f}}F^{\o{b}\u{de}}\partial^{\o{e}}F^{\o{f}}{}_{\u{de}}\partial_{\o{b}}F_{\o{ef}\u{f}}\,.
\label{eq:R11-2}
\end{align}
Terms of the same projection, but with a contraction between the derivatives:
\begin{align}
    &\left[\mathcal{R}^{(1,1)}_{\cancel{\Phi}}\right]^{(3)}_{3}
    =
\frac{1}{4}
\partial^{\u{c}}\partial^{\u{h}}\partial_{\u{c}}F^{\o{c}\u{fg}}F^{\o{d}}{}_{\u{fg}}F_{\o{cd}\u{h}}
+\frac{3}{4}
\partial^{\u{h}}\partial^{\u{c}}\partial_{\u{c}}F^{\o{c}\u{fg}}F^{\o{d}}{}_{\u{fg}}F_{\o{cd}\u{h}}  
+\frac{3}{4}
\partial^{\u{c}}\partial^{\u{d}}F^{\o{cd}}{}_{\u{d}}\partial_{\u{c}}F_{\o{c}}{}^{\u{gh}}F_{\o{d}\u{gh}}
\nonumber\\
&{}
+\frac{1}{4}
\partial^{\u{d}}\partial^{\u{c}}F^{\o{cd}}{}_{\u{d}}\partial_{\u{c}}F_{\o{c}}{}^{\u{gh}}F_{\o{d}\u{gh}}
+\frac{1}{2}
\partial^{\u{c}}\partial^{\u{f}}F^{\o{b}\u{gh}}\partial_{\u{c}}F_{\o{b}}{}^{\o{d}}{}_{\u{f}}F_{\o{d}\u{gh}}
+
\partial^{\u{b}}\partial_{\u{b}}F^{\o{b}\u{gh}}\partial^{\u{f}}F_{\o{b}}{}^{\o{d}}{}_{\u{f}}F_{\o{d}\u{gh}}
\nonumber
\\
&{}
-\frac{1}{4}
\partial^{\u{c}}\partial^{\u{h}}F^{\o{c}\u{fg}}\partial_{\u{c}}F^{\o{d}}{}_{\u{fg}}F_{\o{cd}\u{h}}
+\frac{1}{4}
\partial^{\u{h}}\partial^{\u{c}}F^{\o{c}\u{fg}}\partial_{\u{c}}F^{\o{d}}{}_{\u{fg}}F_{\o{cd}\u{h}}
+
\partial^{\u{b}}\partial_{\u{b}}F^{\o{c}\u{fg}}\partial^{\u{h}}F^{\o{d}}{}_{\u{fg}}F_{\o{cd}\u{h}}
\nonumber
\\
&{}
-\frac{1}{2}
\partial^{\o{b}}\partial^{\u{d}}F^{\o{d}\u{ef}}\partial_{\o{b}}F_{\o{d}}{}^{\o{f}}{}_{\u{d}}F_{\o{f}\u{ef}}
-\frac{1}{2}
\partial^{\u{c}}F^{\o{c}\u{fg}}\partial^{\u{h}}F^{\o{d}}{}_{\u{fg}}\partial_{\u{c}}F_{\o{cd}\u{h}}
-\frac{1}{2}
\partial^{\u{f}}F^{\o{e}\u{de}}\partial^{\o{b}}F^{\o{f}}{}_{\u{de}}\partial_{\o{b}}F_{\o{ef}\u{f}}\,.
\label{eq:R11-3}
\end{align}

Using identities from section \ref{sec:DFT} one finds that
\begin{align}
&\left[\mathcal R^{(1,1)}_{\cancel{\Phi}}\right]^{(3)}_{1}\sim-\frac{2}{3}\mathcal R_{\o a}{}^{\o b\u{de}}\mathcal R_{\o{bc}\u{ef}}\mathcal R^{\o{ca}\u f}{}_{\u d}+\mathcal O(F^4)\,,\\
&\left[\mathcal R^{(1,1)}_{\cancel{\Phi}}\right]^{(3)}_{2}\sim\frac{1}{2}\mathcal R_{\o a}{}^{\o b\u{de}}\mathcal R_{\o{bc}\u{ef}}\mathcal R^{\o{ca}\u f}{}_{\u d}+\mathcal O(F^4)\,,\\
&\left[\mathcal R^{(1,1)}_{\cancel{\Phi}}\right]^{(3)}_{3}\sim\mathcal O(F^4)\,,
\end{align}
so that finally
\begin{equation}
\mathcal R^{(1,1)}=\left[\mathcal R^{(1,1)}_{\cancel{\Phi}}\right]^{(3)}_{1}+\left[\mathcal R^{(1,1)}_{\cancel{\Phi}}\right]^{(3)}_{2}+\left[\mathcal R^{(1,1)}_{\cancel{\Phi}}\right]^{(3)}_{3}+\mathcal O(F^4)
\sim-\frac{1}{6}\mathcal R_{\o a}{}^{\o b\u{de}}\mathcal R_{\o{bc}\u{ef}}\mathcal R^{\o{ca}\u f}{}_{\u d}+\mathcal O(F^4)\,.
\end{equation}
Recall that due to the difference in conventions compared to \cite{Baron:2020xel} the expression in our conventions should be twice this.

\section{Conclusions}
We have shown that the $\alpha'^2$ contributions to the DFT action (\ref{eq:S-DFT}) computed in \cite{Baron:2020xel} simplify to (in our conventions)
\begin{equation}
\mathcal{R}^{(0,2)}\sim \mathcal{R}^{(2,0)}\sim\mathcal O(F^4)\,,\qquad \mathcal{R}^{(1,1)}\sim-\frac13\mathcal R_{\o a}{}^{\o b\u{de}}\mathcal R_{\o{bc}\u{ef}}\mathcal R^{\o{ca}\u f}{}_{\u d}+\mathcal O(F^4)\,.
\end{equation}
Going to the usual supergravity description using (\ref{eq:R-}) and retaining only terms involving the Riemann tensor we get\footnote{To get the signs right one must take into account the extra signs coming from the lower right block of $\eta^{AB}$ in (\ref{eq:eta}).}
\begin{equation}
L=e^{-2\Phi}\left(R-\frac{a+b}{8}R^{abcd}R_{abcd}+\frac{ab}{24}R_a{}^{bde}R_{bcef}R^{caf}{}_d+\ldots\right)\,.
\end{equation}
This agrees precisely with the known expressions \cite{Metsaev:1986yb} up to cubic order in fields, both for the bosonic string ($a=b=-\alpha'$) and the heterotic string ($a=-\alpha'$, $b=0$), since the Gauss-Bonnet combination appearing there is a total derivative at this order. But we glossed over an important point here. The supergravity fields $G$ and $B$ are related to those coming from DFT, $\bar G$ and $\bar B$, by non-covariant field redefinitions \cite{Marques:2015vua}. In particular we have $\bar G=G+\alpha'G^{(1)}+\alpha'^2G^{(2)}$ plus higher order terms, where $G^{(1)}$ is quadratic in the spin connection. However, this does not actually affect the result since these extra terms are only there to make the end result covariant. To see explicitly that they go away we note that under a variation of the metric the Riemann tensor changes as
\begin{equation}
\delta R_{ijkl}=-\nabla_k\nabla_{[i}\delta G_{j]l}+ \nabla_l\nabla_{[i}\delta G_{j]k}-\delta G^m{}_{[i}R_{j]mkl}\,.
\end{equation}
It is easy to see from this that since $G^{(1)}$ is quadratic in fields there are no $\alpha'^2$-terms cubic in fields generated from terms with two $G^{(1)}$'s in the lowest order action. This leaves the terms coming from the $G^{(2)}$ correction in the lowest order action and those coming from the $G^{(1)}$ correction in the order $\alpha'$ action. Since we ignore terms of fourth order or higher the latter are just
\begin{equation}
\alpha'\delta R_{ijkl}R^{ijkl}=
-2\alpha'^2\nabla_k\nabla_iG^{(1)}_{jl}R^{ijkl}
+\ldots
\sim
4\alpha'^2\nabla_i\nabla^jG^{(1)}_{jl}R^{il}
+\ldots\,,
\end{equation}
where the ellipsis denotes higher order terms. Since they are proportional to the Ricci tensor these terms can be canceled by a term in $G^{(2)}$ of the form $G^{(2)}_{ij}=\nabla_{(i}\nabla^kG^{(1)}_{j)k}$. In fact, such a term must be present in $G^{(2)}$ since otherwise the action at order $\alpha'^2$ would not be Lorentz invariant.

There is therefore little doubt that the expressions found in \cite{Baron:2020xel} will reproduce the full $\alpha'^2$-corrections. In fact the latter has been shown to be uniquely fixed by requiring invariance under T-duality on a circle \cite{Garousi:2019mca}. However, to show this will require an enormous amount of work due to the complicated form of the expressions in \cite{Baron:2020xel}. However, as we have seen here, these expressions are highly redundant and we believe there should exist much simpler expressions for the $\alpha'^2$-correction to the DFT action. We plan to report on this in the near future.

\section*{Acknowledgements}
This work is supported by the grant ``Integrable Deformations'' (GA20-04800S) from the Czech Science Foundation (GA\v CR).

\vspace{2cm}
\appendix

\section{Calculation}
Here we give the details of the calculations described in section \ref{sec:calculation}. Throughout this section we use ``$\sim$'' to mean equality up to total derivative terms, equation of motion terms and terms quartic and higher in fields.

\subsection{\texorpdfstring{$\mathcal R^{(0,2)}$}{R(0,2)} term}
We start with showing how to eliminate $\mathcal O(F^2)$ contributions from $\left[\mathcal{R}^{(0,2)}_{\cancel{\Phi}}\right]^{(2)}$ in (\ref{R2}). Using the section condition we combine $ -\frac{1}{2}\partial^{\underline{c}}\partial^{\underline{d}}F^{\overline{b}\underline{gh}}\partial_{\underline{c}}\partial_{\underline{d}}F_{\overline{b}\underline{gh}}+\frac{1}{2}\partial^{\overline{b}}\partial^{\underline{b}}F^{\overline{d}\underline{ef}}\partial_{\overline{b}}\partial_{\underline{b}}F_{\overline{d}\underline{ef}}$ and then integrate by parts and get
\begin{equation}
\partial^{\underline{d}}F^{\overline{b}\underline{gh}}\partial^{\underline{c}}\partial_{\underline{c}}\partial_{\underline{d}}F_{\overline{b}\underline{gh}}
-
\partial^{\underline{d}}F^{\overline{b}\underline{gh}}F^{\underline{c}}\partial_{\underline{c}}\partial_{\underline{d}}F_{\overline{b}\underline{gh}}
\end{equation}
The first term will combine with other terms from (\ref{R2}) and the second we keep for $\mathcal{O}(F^3)$ terms. We have three terms left from (\ref{R2}) 
\begin{align}
&\quad\partial^{\overline{c}}\partial^{\underline{b}}F_{\overline{c}}{}^{\underline{ef}}\partial^{\overline{d}}\partial_{\underline{b}}F_{\overline{d}\underline{ef}}
+
\frac{1}{2}\partial^{\underline{b}}F^{\overline{b}\underline{ef}}\partial^{\overline{d}}\partial_{\overline{b}}\partial_{\underline{b}}F_{\overline{d}\underline{ef}}
+
\frac{3}{2}\partial^{\underline{b}}F^{\overline{b}\underline{ef}}\partial_{\overline{b}}\partial^{\overline{d}}\partial_{\underline{b}}F_{\overline{d}\underline{ef}}
\nonumber\\
\sim&\,
\partial^{\underline{b}}F_{\overline{c}}{}^{\underline{ef}}F^{\overline{c}}\partial^{\overline{d}}\partial_{\underline{b}}F_{\overline{d}\underline{ef}}
+
\partial^{\underline{b}}F^{\overline{b}\underline{ef}}\left(\partial^{\overline{d}}\partial_{\overline{b}}
+
\frac{1}{2}F_{\o{b}}{}^{\overline{d}A}\partial_{A}\right)\partial_{\underline{b}}F_{\overline{d}\underline{ef}}\,.
\end{align}
We rename dummy indices and we have now only two terms left at order $\mathcal{O}(F^2)$,
\begin{equation}
\partial^{\underline{b}}F^{\overline{b}\underline{ef}}\partial^{\underline{c}}\partial_{\underline{c}}\partial_{\underline{b}}F_{\overline{b}\underline{ef}}
+\partial^{\underline{b}}F^{\overline{b}\underline{ef}}\partial^{\overline{d}}\partial_{\overline{b}}
\partial_{\underline{b}}F_{\overline{d}\underline{ef}}\,.
\end{equation}
We commute derivatives and use the Bianchi identity (\ref{eq:Bianchi}) in the second term to get
\begin{equation}
\partial^{\underline{b}}F^{\overline{b}\underline{ef}}\partial^{\underline{c}}\partial_{\underline{c}}\partial_{\underline{b}}F_{\overline{b}\underline{ef}}
+
\partial^{\underline{b}}F^{\overline{b}\underline{ef}}\partial^{\overline{d}}
\partial_{\overline{d}}\partial_{\underline{b}}
F_{\overline{b}\underline{ef}}
+
\partial^{\underline{b}}F^{\overline{b}\underline{ef}}\partial^{\overline{d}}\partial_{\underline{b}}\mathcal R_{\o{bd}\u{ef}}
=
\partial^{\underline{b}}F^{\overline{b}\underline{ef}}\partial^{\overline{d}}\partial_{\underline{b}}\mathcal R_{\o{bd}\u{ef}}\,,
\end{equation}
by using the section condition. The last remaining term can be shown to be zero at leading order as follows
\begin{equation}
\partial_{\underline{b}}\partial^{\overline{d}}\mathcal{R}_{\o{bd}\u{ef}}
=
-2\partial_{\underline{b}}\partial^{\overline{d}}\partial_{[\underline{e}}F_{\underline{f}]\overline{bd}}
=
-2\partial_{\underline{b}}\partial_{[\underline{e}}
\left(
\partial_{\underline{f}]}F_{\overline{b}}
+
\partial^{\overline{d}}
F_{\underline{f}]\overline{bd}}
\right)
+
2\partial_{\underline{b}}\partial_{[\underline{e}}\partial_{\underline{f}]}F_{\overline{b}}
-
2\partial_{\underline{b}}\left(
F^{\o{d}}{}_{\u{e}}{}^{A}\partial_AF_{\underline{f}\overline{bd}}\right)\,.
\end{equation}
Collecting all the $\mathcal{O}(F^3)$ terms generated so far we get the expression in (\ref{R201}). We now want to simplify this and then compare with other terms of order $\mathcal{O}(F^3)$. First we can notice that 
\begin{equation}
3\partial^{\underline{b}}F^{\overline{b}\underline{ef}}\partial^{\overline{d}}\partial_{\underline{b}}\left(F_{[\overline{bd}}{}^EF_{\underline{ef}]E}\right)
\sim
3\partial^{\overline{d}}\partial_{\underline{b}}\partial^{\underline{b}}F^{\overline{b}\underline{ef}}\left(F_{[\overline{bd}}{}^EF_{\underline{ef}]E}\right)
\sim
\frac{3}{2}
\partial_{\underline{b}}\partial^{\underline{b}}\mathcal R^{\o{db}\u{ef}}\left(F_{[\overline{bd}}{}^EF_{\underline{ef}]E}\right) \sim 0\,,
\end{equation}
due to the fact that $\partial^2\mathcal R \sim 0$, (\ref{cur}). Then we use equations of motion to eliminate all the terms with $F_A$, these are 
\begin{equation}
-\partial^{\underline{d}}F^{\overline{b}\underline{gh}}F^{\underline{c}}\partial_{\underline{c}}\partial_{\underline{d}}F_{\overline{b}\underline{gh}}
+
\partial^{\underline{b}}F_{\overline{b}}{}^{\underline{ef}}F^{\overline{b}}\partial^{\overline{d}}\partial_{\underline{b}}F_{\overline{d}\underline{ef}}
+
\partial^{\underline{b}}F^{\overline{b}\underline{ef}}\partial_{\underline{b}}\left(F_{\underline{ef}A}\partial^AF_{\overline{b}}\right)
-2
\partial^{\underline{b}}F^{\overline{b}\underline{ef}}\partial_{\underline{b}}\partial_{\underline{e}}
\left(F^{\o{c}}F_{\u{f}\o{bc}}\right)\,.
\end{equation}
The first term is zero after integration by parts and using the equations of motion $\partial_{\u{c}}F^{\u{c}}=\mathcal O(F^2)$. The other terms can be put into the following form
\begin{equation}
\partial_{\o{d}}F^{\o{d}\underline{ef}}\partial^{\o{b}}F_{\underline{efa}}\partial^{\o{c}}F^{\u{a}}{}_{\o{bc}}
+2\partial^{\o{d}}\partial^{\o{b}}F_{\o{d}}{}^{\u{ef}}\partial^{\o{a}}F_{\u{e}}{}^{\o{c}}{}_{\o{a}}F_{\u{f}\o{bc}}\,,
\end{equation}
using the identities (\ref{eq:Bianchi}), (\ref{eq:eom}) and (\ref{id}).
Now we have eliminated all $F_A$ terms. We split the remaining terms into types depending on their projections.
First we can show that the terms with projections $\partial^{(+)}\partial^{(-)}\partial^{(-)}F^{(-)}F^{(-)}F^{(-)}$ vanish, these terms are
\begin{align}
&{}\partial^{\underline{b}}F^{\overline{b}\underline{ef}}\partial^{\overline{d}}\left(F_{\o{b}\u{b}}{}^{\u{a}}\partial_{\u{a}}F_{\overline{d}\underline{ef}}\right)
+
\partial^{\underline{b}}F^{\overline{b}\underline{ef}}\partial^{\overline{d}}\left(F_{\underline{b}\overline{d}\u{a}}\partial^{\u{a}}F_{\overline{b}\underline{ef}}\right)
+
\partial^{\underline{b}}F^{\overline{b}\underline{ef}}F^{\overline{d}}{}_{\underline{b}}{}^{\u{a}}\partial_{\u{a}}\mathcal R_{\o{bd}\u{ef}}
\nonumber\\
\sim&\,
-\partial^{\overline{d}}\partial^{\underline{b}}F^{\overline{b}\underline{ef}}F_{\o{b}\u{b}}{}^{\u{a}}\partial_{\u{a}}F_{\overline{d}\underline{ef}}
+
\partial^{\underline{b}}F^{\overline{b}\underline{ef}}F_{\underline{b}\overline{d}\u{a}}\partial^{\overline{d}}\partial^{\u{a}}F_{\overline{b}\underline{ef}}
+
\partial^{\underline{b}}F^{\overline{b}\underline{ef}}F^{\overline{d}}{}_{\underline{b}}{}^{\u{a}}\partial_{\u{a}}\mathcal R_{\o{bd}\u{ef}}
\nonumber\\
\sim&\,
\partial^{\u{b}}F^{\o{b}\u{ef}}F_{\u{b}\o{d}\u{a}}\partial^{\u{a}}\mathcal R_{\o{db}\u{ef}}
+
\partial^{\underline{b}}F^{\overline{b}\underline{ef}}F^{\overline{d}}{}_{\underline{b}}{}^{\u{a}}\partial_{\u{a}}\mathcal R_{\o{bd}\u{ef}}
\sim
0\,.
\end{align}
The terms with projection $\partial^{(+)}\partial^{(+)}\partial^{(+)}F^{(-)}F^{(+)}F^{(+)}$ also vanish,
\begin{align}
&{}2\partial^{\underline{b}}F^{\overline{b}\underline{ef}}\partial_{\underline{b}}\left(F^{\overline{d}}{}_{[\underline{f}}{}^{\o{a}}\partial_{\o{a}}F_{\underline{e}]\overline{bd}}\right)
+2\partial^{\o{d}}\partial^{\o{b}}F_{\o{d}}{}^{\u{ef}}\partial^{\o{a}}F_{\u{e}}{}^{\o{c}}{}_{\o{a}}F_{\u{f}\o{bc}}
\nonumber\\
\sim&\,
2\partial^{\o{d}}\partial^{\o{b}}F_{\o{d}}{}^{\u{ef}}F_{\u{e}}{}^{\o{ca}}\partial_{\o{a}}F_{\underline{f}\overline{bc}}
+2\partial^{\o{d}}\partial^{\o{b}}F_{\o{d}}{}^{\u{ef}}\partial^{\o{a}}F_{\u{e}}{}^{\o{c}}{}_{\o{a}}F_{\u{f}\o{bc}}
\sim
-
2\partial^{\o{d}}\partial_{\o{a}}\partial^{\o{b}}F_{\o{d}}{}^{\u{ef}}F_{\u{e}}{}^{\o{ca}}F_{\underline{f}\overline{bc}}\sim0\,.
\end{align}
Now we have 9 terms left, but these are not as easy to simplify,
\begin{align}
&\partial^{\u{b}}F^{\o{b}\u{ef}}F_{\o{b}}{}^{\o{da}}\partial_{\o{a}}\partial_{\u{b}}F_{\o{d}\u{ef}}
+
\frac{1}{2}\partial^{\underline{b}}F^{\overline{b}\underline{ef}}F_{\o{b}}{}^{\o{d}\u{a}}\partial_{\u{a}}\partial_{\underline{b}}F_{\overline{d}\underline{ef}}
+
\partial^{\underline{b}}F^{\overline{b}\underline{ef}}\partial^{\overline{d}}\left(F_{\o{b}\u{b}}{}^{\o{a}}\partial_{\o{a}}F_{\overline{d}\underline{ef}}\right)
\nonumber\\
&{}
+\partial^{\underline{b}}F^{\overline{b}\underline{ef}}\partial^{\overline{d}}\left(F_{\underline{b}\overline{da}}\partial^{\o{a}}F_{\overline{b}\underline{ef}}\right)
+\partial^{\underline{b}}F^{\overline{b}\underline{ef}}F^{\overline{d}}{}_{\underline{b}}{}^{\o{a}}\partial_{\o{a}}\mathcal R_{\o{bd}\u{ef}}
+2\partial^{\underline{b}}F^{\overline{b}\underline{ef}}\partial_{\underline{b}}\left(F^{\overline{d}}{}_{[\underline{f}}{}^{\u{a}}\partial_{\u{a}}F_{\underline{e}]\overline{bd}}\right)
\nonumber\\
&{}
+2\partial^{\underline{b}}F^{\overline{b}\underline{ef}}\partial_{\underline{b}}\partial_{\underline{f}}\left(F_{\o{c}\u{de}}F^{\underline{d}\o{c}}{}_{\o{b}}\right)
+\partial_{\o{d}}F^{\o{d}\underline{ef}}\partial^{\o{b}}F_{\underline{efa}}\partial^{\o{c}}F^{\u{a}}{}_{\o{bc}}
+\frac{1}{2}\partial^{\u{b}}F^{\o{b}\u{ef}}F_{\o{b}}{}^{\o{da}}\partial_{\o{a}}\partial_{\u{b}}F_{\o{d}\u{ef}}\,.
\end{align}
Using Bianchi identities and integration by parts we can combine and simplify these into the following form
\begin{align}
&{}
-F^{\o{b}\underline{ef}}\partial_{\u{a}}\mathcal R^{\o{ad}}{}_{\u{ef}}\partial_{\o{b}}F^{\u{a}}{}_{\o{da}}
+\frac{1}{2}\partial^{\underline{b}}F^{\overline{b}\underline{ef}}F_{\o{b}}{}^{\o{d}\u{a}}\partial_{\u{a}}\partial_{\underline{b}}F_{\overline{d}\underline{ef}}
-\partial^{\underline{b}}\mathcal R^{\o{db}\u{ef}}F_{\o{b}\u{b}}{}^{\o{a}}\mathcal R_{\o{ad}\u{ef}}
+\frac{1}{2}F^{\o{b}\underline{ef}}\partial_{\o{b}}\mathcal R^{\o{ad}}{}_{\u{ef}}\partial_{\u{a}}F^{\u{a}}{}_{\o{da}}
\nonumber\\
&{}
+\partial_{\o{d}}F^{\o{d}\underline{ef}}\partial_{\u{a}}F_{\o{b}\u{ef}}\partial^{\o{c}}F^{\u{a}}{}_{\o{bc}}
-\frac{1}{2}\partial^{\u{b}}F^{\o{b}\u{ef}}F_{\o{b}}{}^{\o{da}}\partial_{\o{a}}\partial_{\u{b}}F_{\o{d}\u{ef}}\,.
    \label{OF2}
\end{align}
Now we turn to the $\mathcal O(F^3)$ terms and then combine the result with these six terms we have left. The $\left[\mathcal{R}^{(0,2)}_{\Phi}\right]^{(3)}$ terms in (\ref{R3withf}) are quite simple,
\begin{align}
&{}-2\partial^{\u{b}}F^{\o{d}\u{ef}}\partial^{\o{c}}\partial_{\u{b}}F_{\o{c}\u{ef}}F_{\o{d}}
-2\partial^{\u{b}}F^{\o{c}\u{ef}}\partial_{\o{c}}\partial_{\u{b}}F^{\o{d}}{}_{\u{ef}}F_{\o{d}}
-2\partial^{\u{b}}F^{\o{c}\u{ef}}\partial_{\u{b}}F^{\o{d}}{}_{\u{ef}}\partial_{\o{c}}F_{\o{d}}
\nonumber\\
\sim&\,
2\partial^{\u{b}}F^{\o{c}\u{ef}}\partial_{\u{b}}F^{\o{d}}{}_{\u{ef}}\partial_{\o{c}}F_{\o{d}}
-2\partial^{\u{b}}F^{\o{c}\u{ef}}\partial_{\u{b}}F^{\o{d}}{}_{\u{ef}}\partial_{\o{c}}F_{\o{d}}
\sim0\,.
\end{align}
We have split the $\left[\mathcal{R}^{(0,2)}_{\cancel{\Phi}}\right]^{(3)}$ fields into four. First we show that the $\left[\mathcal{R}^{(0,2)}_{\cancel{\Phi}}\right]^{(3)}_{3}$ terms in (\ref{R33}) actually vanish. The first two cancel after commutation of the derivatives. The other three are
\begin{align}
&{}2
\partial^{\u{g}}F^{\o{a}\u{ef}}\partial^{\o{d}}\partial^{\u{h}}F_{\o{a}\u{ef}}F_{\o{d}\u{gh}}
-2
\partial^{\u{g}}F^{\o{a}\u{ef}}\partial_{\o{a}}\partial^{\u{h}}F^{\o{d}}{}_{\u{ef}}F_{\o{d}\u{gh}}
+2
\partial^{\u{g}}F^{\o{a}\u{ef}}\partial^{\u{h}}F^{\o{d}}{}_{\u{ef}}\partial_{\o{a}}F_{\o{d}\u{gh}}
\nonumber\\
\sim&\,
2
\partial^{\u{g}}F^{\o{a}\u{ef}}\partial^{\o{d}}\partial^{\u{h}}F_{\o{a}\u{ef}}F_{\o{d}\u{gh}}
-2
\partial^{\u{g}}F^{\o{a}\u{ef}}\partial_{\o{a}}\partial^{\u{h}}F^{\o{d}}{}_{\u{ef}}F_{\o{d}\u{gh}}
\sim
\partial^{\u{g}}F^{\o{a}\u{ef}}\partial^{\u{h}}\mathcal R_{\o{da}\u{ef}}F^{\o{d}}{}_{\u{gh}}
\\
\sim&\,
-
\partial^{\u{g}}F^{\o{a}\u{ef}}\mathcal R_{\o{da}\u{ef}}\partial^{\u{h}}F^{\o{d}}{}_{\u{gh}}
\sim
\partial^{\u{g}}F^{\o{a}\u{ef}}\mathcal R_{\o{da}\u{ef}}\partial^{\o{d}}F_{\u{g}}
\sim
-
\partial^{\u{g}}\partial^{\o{d}}F^{\o{a}\u{ef}}\mathcal R_{\o{da}\u{ef}}F_{\u{g}}
\sim
\frac{1}{2}\partial^{\u{g}}\mathcal R^{\o{da}\u{ef}}\mathcal R_{\o{da}\u{ef}}F_{\u{g}}\sim 0\,.
\nonumber
\end{align}
The last step gives zero via integration by parts and using the equations of motion $\partial^{\u{g}}F_{\u{g}}=\mathcal O(F^2)$.
Using integration by parts and commutation of derivatives we can reduce the $\left[\mathcal{R}^{(0,2)}_{\cancel{\Phi}}\right]^{(3)}_{1}$ terms in (\ref{R31}) down to three
\begin{align}
&{}2\partial^{\o{c}}\partial_{\o{c}}\partial^{\u{f}}F^{\o{f}\u{de}}F_{\o{f}}{}^{\o{e}}{}_{\u{f}}F_{\o{e}\u{de}}
-
2\partial^{\u{f}}\partial^{\o{c}}\partial_{\o{b}}F^{\o{b}\u{de}}F_{\o{e}\u{de}}F_{\o{ce}\u{f}}
+\frac{3}{2}\partial^{\u{h}}\partial^{\u{c}}F^{\o{c}\u{fg}}\partial_{\u{c}}F^{\o{d}}{}_{\u{fg}}F_{\o{cd}\u{h}}
\\
\sim&\,
2F_{\o{e}\u{de}}F_{\o{ce}\u{f}}
\left(
\partial_{\o{b}}\partial^{\u{f}}\partial^{\o{b}}F^{\o{c}\u{de}}-\partial_{\o{b}}\partial^{\u{f}}\partial^{\o{c}}F^{\o{b}\u{de}}
\right)
+\frac{3}{2}\partial^{\u{h}}\partial^{\u{c}}F^{\o{c}\u{fg}}\partial_{\u{c}}F^{\o{d}}{}_{\u{fg}}F_{\o{cd}\u{h}}
\sim
\frac{3}{2}\partial^{\u{h}}\partial^{\u{c}}F^{\o{c}\u{fg}}\partial_{\u{c}}F^{\o{d}}{}_{\u{fg}}F_{\o{cd}\u{h}}\,.
\nonumber
\end{align}
Analogously the $\left[\mathcal{R}^{(0,2)}_{\cancel{\Phi}}\right]^{(3)}_{2}$ terms in (\ref{R32}) become, after using integration by parts and commutation of derivatives,
\begin{align}
&{}
8\partial^{\o{c}}\partial^{\o{f}}F^{\o{e}\u{ce}}\partial_{\o{c}}F_{\o{e}\u{c}}{}^{\u{f}}F_{\o{f}\u{ef}}
-16
\partial^{\o{c}}\partial^{\o{e}}F_{\o{f}\u{ce}}\partial_{\o{c}}F_{\o{e}\u{c}}{}^{\u{f}}F_{\o{f}\u{ef}}
+
12\partial^{\o{c}}\partial^{\o{d}}F_{\o{d}}{}^{\u{ce}}\partial_{\o{c}}F^{\o{f}}{}_{\u{c}}{}^{\u{f}}F_{\o{f}\u{ef}}
-8
\partial^{\o{d}}\partial^{\o{e}}F_{\o{d}}{}^{\u{ce}}\partial^{\o{f}}F_{\o{e}\u{c}}{}^{\u{f}}F_{\o{f}\u{ef}}
\nonumber\\
\sim&\,
-8
\partial^{\o{c}}\partial^{\o{e}}F_{\o{f}\u{ce}}\partial_{\o{c}}F_{\o{e}\u{c}}{}^{\u{f}}F_{\o{f}\u{ef}}
+4
\partial^{\o{c}}\partial^{\o{d}}F_{\o{d}}{}^{\u{ce}}\partial_{\o{c}}F^{\o{f}}{}_{\u{c}}{}^{\u{f}}F_{\o{f}\u{ef}}
+4
\partial^{\o{d}}\partial^{\o{f}}F_{\o{d}}{}^{\u{ce}}\partial^{\o{e}}F_{\o{e}\u{c}}{}^{\u{f}}F_{\o{f}\u{ef}}
\nonumber\\
\sim&\,
8
\partial^{\o{e}}F_{\o{f}\u{ce}}\partial^{\o{c}}\partial_{\o{c}}F_{\o{e}\u{c}}{}^{\u{f}}F_{\o{f}\u{ef}}
+4
\partial^{\o{c}}\partial^{\o{d}}F_{\o{d}}{}^{\u{ce}}\partial_{\o{c}}F^{\o{f}}{}_{\u{c}}{}^{\u{f}}F_{\o{f}\u{ef}}
+4
\partial^{\o{d}}\partial^{\o{f}}F_{\o{d}}{}^{\u{ce}}\partial^{\o{e}}F_{\o{e}\u{c}}{}^{\u{f}}F_{\o{f}\u{ef}}
\nonumber\\
\sim&\,
-4
\partial^{\o{c}}\partial^{\o{d}}F_{\o{d}}{}^{\u{ce}}\partial_{\o{c}}F^{\o{f}}{}_{\u{c}}{}^{\u{f}}F_{\o{f}\u{ef}}
+4
\partial^{\o{d}}\partial^{\o{f}}F_{\o{d}}{}^{\u{ce}}\partial^{\o{e}}F_{\o{e}\u{c}}{}^{\u{f}}F_{\o{f}\u{ef}}
\nonumber\\
\sim&\,
-4
\partial^{\o{d}}F_{\o{d}}{}^{\u{ce}}\partial^{\o{e}}\partial^{\o{f}}F_{\o{e}\u{c}}{}^{\u{f}}F_{\o{f}\u{ef}}
-4
\partial^{\o{c}}\partial^{\o{d}}F_{\o{d}}{}^{\u{ce}}\partial_{\o{c}}F^{\o{f}}{}_{\u{c}}{}^{\u{f}}F_{\o{f}\u{ef}}
\nonumber\\
\sim&\,
-4
\partial^{\o{d}}F_{\o{d}}{}^{\u{ce}}\partial^{\o{e}}\partial_{\o{e}}F^{\o{f}}{}_{\u{c}}{}^{\u{f}}F_{\o{f}\u{ef}}
-4
\partial^{\o{e}}\partial^{\o{d}}F_{\o{d}}{}^{\u{ce}}\partial_{\o{e}}F^{\o{f}}{}_{\u{c}}{}^{\u{f}}F_{\o{f}\u{ef}}
\sim
4
\partial^{\o{d}}F_{\o{d}}{}^{\u{ce}}\partial^{\o{e}}F^{\o{f}}{}_{\u{c}}{}^{\u{f}}\partial_{\o{e}}F_{\o{f}\u{ef}}\sim0\,.
\end{align}
Having simplified these let us combine them with the terms (\ref{OF2}) from the previous calculation. 
We have six terms left from $\mathcal{O}(F^2)$ and just two terms from $\left[\mathcal{R}^{(0,2)}_{\cancel{\Phi}}\right]^{(3)}$. We have
\begin{align}
&{}
-F^{\o{b}\underline{ef}}\partial_{\u{a}}\mathcal R^{\o{ad}}{}_{\u{ef}}\partial_{\o{b}}F^{\u{a}}{}_{\o{da}}
+\frac{1}{2}\partial^{\underline{b}}F^{\overline{b}\underline{ef}}F_{\o{b}}{}^{\o{d}\u{a}}\partial_{\u{a}}\partial_{\underline{b}}F_{\overline{d}\underline{ef}}
-\partial^{\underline{b}}\mathcal R^{\o{db}\u{ef}}F_{\o{b}\u{b}}{}^{\o{a}}\mathcal R_{\o{ad}\u{ef}}
+\frac{1}{2}F^{\o{b}\underline{ef}}\partial_{\o{b}}\mathcal R^{\o{ad}}{}_{\u{ef}}\partial_{\u{a}}F^{\u{a}}{}_{\o{da}}
\nonumber\\
&{}
+\partial_{\o{d}}F^{\o{d}\underline{ef}}\partial_{\u{a}}F_{\o{b}\u{ef}}\partial^{\o{c}}F^{\u{a}}{}_{\o{bc}}
+\frac{3}{2}\partial^{\u{h}}\partial^{\u{c}}F^{\o{c}\u{fg}}\partial_{\u{c}}F^{\o{d}}{}_{\u{fg}}F_{\o{cd}\u{h}}
\sim
-\frac{1}{2}F^{\u{a}}{}_{\o{da}}\partial_{\o{b}}F^{\o{b}\underline{ef}}\partial_{\u{a}}\mathcal R^{\o{ad}}{}_{\u{ef}}
\nonumber\\
&{}
+\frac{1}{2}F^{\o{b}\underline{ef}}\partial_{\o{b}}\mathcal R^{\o{ad}}{}_{\u{ef}}\partial_{\u{a}}F^{\u{a}}{}_{\o{da}}
-F^{\o{b}\u{ef}}\partial_{\u{a}}\mathcal R^{\o{ad}}{}_{\u{ef}}\partial_{\o{b}}F^{\u{a}}{}_{\o{da}}
-\partial^{\underline{b}}\mathcal R^{\o{db}\u{ef}}F_{\o{b}\u{b}}{}^{\o{a}}\mathcal R_{\o{ad}\u{ef}}
\sim
-\partial^{\underline{b}}\mathcal R^{\o{db}\u{ef}}F_{\o{b}\u{b}}{}^{\o{a}}\mathcal R_{\o{ad}\u{ef}}
\nonumber\\
&{}
-\frac{1}{2}\partial_{\u{a}}F^{\o{b}\underline{ef}}\partial_{\o{b}}\mathcal R^{\o{ad}}{}_{\u{ef}}F^{\u{a}}{}_{\o{da}}
-\frac{1}{2}F^{\o{b}\u{ef}}\partial_{\u{a}}\mathcal R^{\o{ad}}{}_{\u{ef}}\partial_{\o{b}}F^{\u{a}}{}_{\o{da}}
\sim
\partial^{\underline{b}}\mathcal R^{\o{db}\u{ef}}\partial_{\o{a}}F_{\o{b}\u{b}}{}^{\o{a}}F_{\o{d}\u{ef}}
-\partial^{\underline{b}}\mathcal R^{\o{db}\u{ef}}\partial_{\o{d}}F_{\o{b}\u{b}}{}^{\o{a}}F_{\o{a}\u{ef}}
\nonumber\\
&{}
+F^{\o{b}\underline{ef}}\partial^{\o{a}}\mathcal R^{\o{bd}}{}_{\u{ef}}\partial_{\u{a}}F^{\u{a}}{}_{\o{da}}
-\frac{1}{2}F^{\o{b}\u{ef}}\partial_{\u{a}}\mathcal R^{\o{ad}}{}_{\u{ef}}\partial_{\o{b}}F^{\u{a}}{}_{\o{da}}
\sim
\partial^{\underline{b}}\mathcal R^{\o{db}\u{ef}}\partial_{\o{a}}F_{\o{b}\u{b}}{}^{\o{a}}F_{\o{d}\u{ef}}
-\partial^{\underline{b}}\mathcal R^{\o{db}\u{ef}}\partial_{\o{d}}F_{\o{b}\u{b}}{}^{\o{a}}F_{\o{a}\u{ef}}
\nonumber\\
&{}
-F^{\o{b}\underline{ef}}\partial^{\o{a}}\mathcal R^{\o{bd}}{}_{\u{ef}}\left(\partial_{\o{c}}F^{\o{c}}{}_{\o{da}}+2\partial_{[\o{d}}F_{\o{a}]}\right)
-\frac{1}{2}F^{\o{b}\u{ef}}\partial_{\u{a}}\mathcal R^{\o{ad}}{}_{\u{ef}}\partial_{\o{b}}F^{\u{a}}{}_{\o{da}}
\nonumber\\
\sim&\,
\partial^{\underline{b}}\mathcal R^{\o{db}\u{ef}}\partial_{\o{a}}F_{\o{b}\u{b}}{}^{\o{a}}F_{\o{d}\u{ef}}
-F^{\o{b}\underline{ef}}\partial_{\o{a}}\mathcal R^{\o{bd}}{}_{\u{ef}}\partial_{\o{c}}F^{\o{c}}{}_{\o{d}}{}^{\o{a}}
-\frac{1}{2}F^{\o{b}\u{ef}}\partial_{\u{a}}\mathcal R^{\o{ad}}{}_{\u{ef}}\partial_{\o{b}}F^{\u{a}}{}_{\o{da}}
\nonumber\\
\sim&\,
\mathcal R^{\o{cb}\u{ef}}\partial_{\o{a}}\mathcal R^{\o{bd}}{}_{\u{ef}}F_{\o{cd}}{}^{\o{a}}
-F^{\o{b}\u{ef}}\partial_{\u{a}}\mathcal R^{\o{ad}}{}_{\u{ef}}\partial_{\o{d}}F^{\u{a}}{}_{\o{ba}}
-\partial^{\underline{b}}\mathcal R^{\o{db}\u{ef}}\partial_{\o{d}}F_{\o{b}\u{b}}{}^{\o{a}}F_{\o{a}\u{ef}}
\nonumber\\
\sim&\,
\mathcal R^{\o{cb}\u{ef}}\partial_{\o{a}}\mathcal R^{\o{bd}}{}_{\u{ef}}F_{\o{cd}}{}^{\o{a}}
\sim
-\frac{1}{2}\mathcal R^{\o{cb}\u{ef}}\mathcal R^{\o{ad}}{}_{\u{ef}}\partial_{\o{b}}F_{\o{cda}}
\sim0\,.
\end{align}
The last step is due to the Bianchi identity (\ref{eq:Bianchi}). This completes the proof that $\mathcal{R}^{(2,0)}\sim\mathcal{O}(F^4)$.

\subsection{\texorpdfstring{$\mathcal R^{(1,1)}$}{R(1,1)} term}
We start with $\left[\mathcal{R}^{(1,1)}_{\cancel{\Phi}}\right]^{(3)}_{3}$ in (\ref{eq:R11-3}). We can use only commutation of derivatives and integration by parts to simplify these as follows,
\begin{align}
&{}
-\partial^{\u{f}}\partial^{\o{d}}F^{\o{ef}}{}_{\u{f}}\partial_{\o{d}}F_{\o{e}}{}^{\u{de}}F_{\o{f}\u{de}}
-\partial^{\o{d}}\partial^{\u{f}}F^{\o{e}\u{de}}\partial_{\o{d}}F_{\o{e}}{}^{\o{f}}{}_{\u{f}}F_{\o{f}\u{de}}
\sim
\partial^{\o{d}}F^{\o{e}\u{de}}\partial_{\o{d}}F_{\o{e}}{}^{\o{f}}{}_{\u{f}}\partial^{\u{f}}F_{\o{f}\u{de}}
\nonumber\\
\sim&\,
-\partial^{\u{c}}F^{\o{e}\u{de}}\partial_{\u{c}}F_{\o{e}}{}^{\o{f}}{}_{\u{f}}\partial^{\u{f}}F_{\o{f}\u{de}}
\sim
-\frac{1}{2}\partial^{\u{c}}F^{\o{e}\u{de}}\mathcal R_{\u{cf}\o{ef}}\partial^{\u{f}}F^{\o{f}}{}_{\u{de}}\sim0\,.
\end{align}
Next we consider $\left[\mathcal{R}^{(1,1)}_{\cancel{\Phi}}\right]^{(3)}_1$ in (\ref{eq:R11-1}). We have
\begin{align}
&{}
\frac{4}{3}
\partial^{\o{e}}F^{\o{c}\u{ce}}\partial_{\o{c}}F_{\o{f}\u{ef}}\partial^{\o{f}}F_{\o{e}\u{c}}{}^{\u{f}}
-4
\partial^{\o{b}}F^{\o{c}\u{ce}}\partial_{\o{c}}F^{\o{f}}{}_{\u{ef}}\partial_{\o{b}}F_{\o{f}\u{c}}{}^{\u{f}}
\nonumber\\
\sim&\,
\frac{4}{3}
\partial^{\o{e}}F^{\o{c}\u{ce}}\partial_{\o{c}}F_{\o{f}\u{ef}}
\left(
\partial_{\o{e}}F^{\o{f}}{}_{\u{c}}{}^{\u{f}}-\mathcal R_{\o e}{}^{\o{f}}{}_{\u{c}}{}^{\u f}
\right)
-4
\partial^{\o{b}}F^{\o{c}\u{ce}}\partial_{\o{c}}F^{\o{f}}{}_{\u{ef}}\partial_{\o{b}}F_{\o{f}\u{c}}{}^{\u{f}}
\nonumber\\
\sim&\,
-\frac{8}{3}
\partial^{\o{e}}F^{\o{c}\u{ce}}\partial_{\o{c}}F^{\o{f}}{}_{\u{ef}}\partial_{\o{e}}F_{\o{f}\u{c}}{}^{\u{f}}
-
\frac{4}{3}
\partial^{\o{e}}F^{\o{c}\u{ce}}\partial_{\o{c}}F^{\o{f}}{}_{\u{ef}}
\mathcal{R}_{\o{ef}}{}_{\u{c}}{}^{\u f}
\nonumber\\
\sim&\,
\frac{4}{3}\partial^{\o{c}}F^{\o{e}\u{ce}}\mathcal{R}_{\o{ef}\u{cf}}\partial_{\o{c}}F^{\o{f}}{}_{\u{e}}{}^{\u{f}}
-
\frac{4}{3}
\partial^{\o{e}}F^{\o{c}\u{ce}}\partial_{\o{c}}F^{\o{f}}{}_{\u{ef}}
\mathcal{R}_{\o{ef}}{}_{\u{c}}{}^{\u f}
\nonumber\\
\sim&\,
\frac{4}{3}\mathcal{R}_{\o{fe}\u{cf}}\partial_{\o{c}}F^{\o{f}}{}_{\u{e}}{}^{\u{f}}\left(\partial^{\o{e}}F^{\o{c}\u{ce}}-\partial^{\o{c}}F^{\o{e}\u{ce}}\right)
\sim
\frac{2}{3}\mathcal R^{\o{ec}\u{ce}}\mathcal R_{\o{cf}\u{ef}}\mathcal R_{\o f}{}^{\o{e}\u{c}}{}_{\u f}\,.
\end{align}

Lastly the $\left[\mathcal{R}^{(1,1)}_{\cancel{\Phi}}\right]^{(3)}_2$ terms in (\ref{eq:R11-2}) simplify to
\begin{align}
&{}
\partial^{\o{d}}\partial^{\u{f}}F^{\o{ef}}{}_{\u{f}}\partial_{\o{e}}F_{\o{f}}{}^{\u{de}}F_{\o{d}\u{de}}
+
\frac{1}{2}
\partial^{\o{d}}F^{\o{ef}}{}_{\u{f}}\partial^{\u{f}}\partial_{\o{e}}F_{\o{f}}{}^{\u{de}}F_{\o{d}\u{de}}
-\frac{1}{2}F_{\o{ef}\u{f}}\partial^{\o{d}}\partial^{\o{e}}F^{\o{f}\u{de}}\partial^{\u{f}}F_{\o{d}\u{de}}
+\partial^{\o{d}}F^{\o{ef}}{}_{\u{f}}\partial_{\o{e}}F_{\o{f}}{}^{\u{de}}\partial^{\u{f}}F_{\o{d}\u{de}}
\nonumber\\
\sim&\,
-\frac{1}{2}
\partial^{\o{d}}F^{\o{ef}}{}_{\u{f}}\partial_{\o{e}}\partial^{\u{f}}F_{\o{f}}{}^{\u{de}}F_{\o{d}\u{de}}
-\frac{1}{2}
F_{\o{ef}\u{f}}\partial^{\o{d}}\partial^{\o{e}}F^{\o{f}\u{de}}\partial^{\u{f}}F_{\o{d}\u{de}}\,.
\end{align}
To show that these two terms are actually proportional to $R^3$ it is simpler to work in the opposite direction, by taking $\mathcal R_{\o a}{}^{\o{b}\u{de}}\mathcal R_{\o{bc}\u{ef}}\mathcal R^{\o{ca}\u f}{}_{\u{d}}$ and showing that it is equal to the two terms at this order of fields. We have
\begin{align}
\frac{1}{2}\mathcal R_{\o a}{}^{\o{b}\u{de}}\mathcal R_{\o{bc}\u{ef}}\mathcal R^{\o{ca}\u{f}}{}_{\u d}
\sim&
\partial_{\o{a}}F_{\o{b}\u{cd}}\mathcal R^{\o{ae}\u{c}}{}_{\u{f}}\mathcal R^{\o{b}}{}_{\o{e}}{}^{\u{df}}
\sim
-F_{\o{b}\u{cd}}\mathcal R^{\o{ae}\u{c}}{}_{\u{f}}\partial_{\o{a}}\mathcal R^{\o{b}}{}_{\o{e}}{}^{\u{df}}
\nonumber\\
\sim&\,
\frac{1}{2}F_{\o{b}\u{cd}}\partial^{\u{c}}F_{\u{f}}{}^{\o{ae}}\partial^{\o{b}}\mathcal R_{\o{ae}}{}^{\u{df}}
-\frac{1}{2}F_{\o{b}\u{cd}}\partial_{\u{f}}F^{\u{c}\o{ae}}\partial^{\o{b}}\mathcal R_{\o{ae}}{}^{\u{df}}\,.
\end{align}
Taking the first term here and integrating by parts we get
\begin{equation}
-\frac{1}{2}\partial^{\u{c}}F_{\o{b}\u{cd}}F_{\u{f}}{}^{\o{ae}}\partial^{\o{b}}\mathcal R_{\o{ae}}{}^{\u{df}}
-\frac{1}{2}F_{\o{b}\u{cd}}F_{\u{f}}{}^{\o{ae}}\partial^{\o{b}}\partial^{\u{c}}\mathcal R_{\o{ae}}{}^{\u{df}}\,.
\end{equation}
The the second term here is
\begin{equation}
\frac{1}{4}F_{\o{b}\u{cd}}F_{\u{f}}{}^{\o{ae}}\partial^{\o{b}}\partial^{\u{f}}\mathcal R_{\o{ae}}{}^{\u{cd}}
\sim
-\frac{1}{4}\partial^{\o{b}}F_{\o{b}\u{cd}}F_{\u{f}}{}^{\o{ae}}\partial^{\u{f}}\mathcal R_{\o{ae}}{}^{\u{cd}}
-\frac{1}{4}F_{\o{b}\u{cd}}\partial^{\o{b}}F_{\u{f}}{}^{\o{ae}}\partial^{\u{f}}\mathcal R_{\o{ae}}{}^{\u{cd}}\,.
\end{equation}
The second term here is precisely one of the two that we wanted, and we have two terms unaccounted for:
\begin{equation}
-\frac{1}{2}\partial^{\u{c}}F_{\o{b}\u{cd}}F_{\u{f}}{}^{\o{ae}}\partial^{\o{b}}\mathcal R_{\o{ae}}{}^{\u{df}}
-\frac{1}{4}\partial^{\o{b}}F_{\o{b}\u{cd}}F_{\u{f}}{}^{\o{ae}}\partial^{\u{f}}\mathcal R_{\o{ae}}{}^{\u{cd}}\,.
\end{equation}
We also have
\begin{equation}
-\frac{1}{2}F_{\o{b}\u{cd}}\partial_{\u{f}}F^{\u{c}\o{ae}}\partial^{\o{b}}\mathcal R_{\o{ae}}{}^{\u{df}}
\sim
\frac{1}{2}\partial_{\u{f}}F_{\o{b}\u{cd}}F^{\u{c}\o{ae}}\partial^{\o{b}}\mathcal R_{\o{ae}}{}^{\u{df}}
\sim
\frac{1}{4}\partial_{\o{b}}F_{\u{fcd}}F^{\u{c}\o{ae}}\partial^{\o{b}}\mathcal R_{\o{ae}}{}^{\u{df}}
-\frac{1}{4}\partial_{\u{c}}F_{\o{b}\u{df}}F^{\u{c}\o{ae}}\partial^{\o{b}}\mathcal R_{\o{ae}}{}^{\u{df}}\,.
\end{equation}
The second term here is the same as the second term we wanted, namely $-\frac{1}{2}F_{\o{ef}\u{f}}\partial^{\o{d}}\partial^{\o{e}}F^{\o{f}\u{de}}\partial^{\u{f}}F_{\o{d}\u{de}}$.
However, we have three terms left over which have to cancel
\begin{equation}
\label{three}
-\frac{1}{2}\partial^{\u{c}}F_{\o{b}\u{cd}}F_{\u{f}}{}^{\o{ae}}\partial^{\o{b}}\mathcal R_{\o{ae}}{}^{\u{df}}
-\frac{1}{4}\partial^{\o{b}}F_{\o{b}\u{cd}}F_{\u{f}}{}^{\o{ae}}\partial^{\u{f}}\mathcal R_{\o{ae}}{}^{\u{cd}}
+\frac{1}{4}\partial_{\o{b}}F_{\u{fcd}}F^{\u{c}\o{ae}}\partial^{\o{b}}\mathcal R_{\o{ae}}{}^{\u{df}}\,.
\end{equation}
The second term becomes after, using the Bianchi identity,
\begin{equation}
-\frac{1}{4}\partial^{\o{b}}F_{\o{b}\u{cd}}F_{\u{f}}{}^{\o{ae}}\partial^{\u{f}}\mathcal R_{\o{ae}}{}^{\u{cd}}
\sim
\frac{1}{4}\partial^{\u{b}}F_{\u{bcd}}F_{\u{f}}{}^{\o{ae}}\partial^{\u{f}}\mathcal R_{\o{ae}}{}^{\u{cd}}
+\frac{1}{2}\partial_{\u{c}}F_{\u{d}}F_{\u{f}}{}^{\o{ae}}\partial^{\u{f}}\mathcal R_{\o{ae}}{}^{\u{cd}}\,.
\end{equation}
The second term here combines with the first term from (\ref{three}),
\begin{align}
&{}
-
\frac{1}{2}\partial^{\u{c}}F_{\o{b}\u{cd}}F_{\u{f}}{}^{\o{ae}}\partial^{\o{b}}\mathcal R_{\o{ae}}{}^{\u{df}}
+
\frac{1}{2}\partial_{\u{c}}F_{\u{d}}F_{\u{f}}{}^{\o{ae}}\partial^{\u{f}}\mathcal R_{\o{ae}}{}^{\u{cd}}
\sim
\frac{1}{2}
\partial_{\u{c}}F_{\u{d}}F_{\u{f}}{}^{\o{ae}}\partial^{\u{c}}\mathcal R_{\o{ae}}{}^{\u{df}}
+
\frac{1}{2}\partial_{\u{c}}F_{\u{d}}F_{\u{f}}{}^{\o{ae}}\partial^{\u{f}}\mathcal R_{\o{ae}}{}^{\u{cd}}
\nonumber\\
\sim&\,
\frac{1}{2}
\partial_{\u{c}}F_{\u{d}}F_{\u{f}}{}^{\o{ae}}\partial^{\u{c}}\partial^{\u{f}}F^{\u{d}}{}_{\o{ae}}
-
\frac{1}{2}
\partial_{\u{c}}F_{\u{d}}F_{\u{f}}{}^{\o{ae}}\partial^{\u{c}}\partial^{\u{d}}F^{\u{f}}{}_{\o{ae}}
-
\frac{1}{2}\partial_{\u{c}}F_{\u{d}}F_{\u{f}}{}^{\o{ae}}\partial^{\u{f}}\partial^{\u{c}}F^{\u{d}}{}_{\o{ae}}
+
\frac{1}{2}\partial_{\u{c}}F_{\u{d}}F_{\u{f}}{}^{\o{ae}}\partial^{\u{f}}\partial^{\u{d}}F^{\u{c}}{}_{\o{ae}}
\nonumber\\
\sim&\,
\frac{1}{2}\partial_{\u{c}}F_{\u{d}}F_{\u{f}}{}^{\o{ae}}\partial^{\u{d}}\mathcal R_{\o{ae}}{}^{\u{cf}}
\sim
-\frac{1}{2}F_{\u{d}}\partial_{\u{c}}F_{\u{f}}{}^{\o{ae}}\partial^{\u{d}}\mathcal R_{\o{ae}}{}^{\u{cf}}
\sim
-\frac{1}{4}F_{\u{d}}\mathcal R_{\u{cf}}{}^{\o{ae}}\partial^{\u{d}}\mathcal R_{\o{ae}}{}^{\u{cf}}\sim 0\,.
\end{align}
And we have two terms which are left,
\begin{align}
&{}
\frac{1}{4}\partial^{\u{b}}F_{\u{bcd}}F_{\u{f}\o{ae}}\partial^{\u{f}}\mathcal R^{\o{ae}\u{cd}}
-\frac{1}{4}\partial^{\u{b}}F_{\u{fcd}}F^{\u{c}}{}_{\o{ae}}\partial_{\u{b}}\mathcal R^{\o{ae}\u{df}}
\sim
-\frac{1}{4}F_{\u{bfd}}\partial^{\u{b}}F_{\u{c}\o{ae}}\partial^{\u{c}}\mathcal R^{\o{ae}\u{fd}}
+\frac{1}{4}F_{\u{fcd}}\partial_{\u{b}}F^{\u{c}}{}_{\o{ae}}\partial^{\u{b}}\mathcal R^{\o{ae}\u{df}}
\nonumber\\
\sim&\,
-\frac{1}{4}F_{\u{bfd}}\partial^{\u{b}}F_{\u{c}\o{ae}}\partial^{\u{c}}\mathcal R^{\o{ae}\u{fd}}
+\frac{1}{4}F_{\u{bfd}}\partial_{\u{c}}F_{\u{b}\o{ae}}\partial^{\u{c}}\mathcal R^{\o{ae}\u{fd}}
\sim
\frac{1}{8}F^{\u{b}}{}_{\u{fd}}\mathcal R_{\u{cb}\o{ae}}\partial^{\u{c}}\mathcal R^{\o{ae}\u{fd}}
\nonumber\\
\sim&\,
-\frac{1}{8}\partial^{\u{c}}F^{\u{b}}{}_{\u{fd}}\mathcal R_{\u{cb}\o{ae}}\mathcal R^{\o{ae}\u{fd}}=0\,,
\end{align}
where we used the Bianchi identity $\partial_{[\u{c}}F_{\u{bfd}]}=\mathcal O(F^2)$.

Putting this together we have shown that
\begin{equation}
\mathcal R^{(1,1)}\sim
-\frac{2}{3}\mathcal R_{\o e}{}^{\o{c}\u{ce}}\mathcal R_{\o{cf}\u{ef}}\mathcal R^{\o{fe}\u{f}}{}_{\u c}
+\frac{1}{2}\mathcal R_{\o a}{}^{\o{b}\u{ce}}\mathcal R_{\o{bc}\u{ef}}\mathcal R^{\o{ca}\u f}{}_{\u{c}}
+\mathcal O(F^4)
=
-\frac{1}{6}\mathcal R_{\o a}{}^{\o b\u{de}}\mathcal R_{\o{bc}\u{ef}}\mathcal R^{\o{ca}\u f}{}_{\u d}
+\mathcal O(F^4)\,.
\end{equation}

\bibliographystyle{nb}
\bibliography{biblio}{}
\end{document}